\documentclass[twocolumn,tighten]{aastex631}

\submitjournal{ApJ}
\shortauthors{Ghosh and Afrin}
\usepackage{graphicx}	
\usepackage{amsmath}	
\usepackage{amsfonts}
\usepackage{float}
\usepackage{multirow}
\usepackage{textcomp}
\usepackage{wrapfig}

\usepackage{eucal}
\hypersetup{
    colorlinks=true,
    citecolor=magenta,
    linkcolor=cyan,
}
\begin{document}
\title{An Upper Limit on the Charge of the Black Hole Sgr A* from EHT Observations}
\author[0000-0002-0835-3690]{Sushant G. Ghosh}
\affiliation{Centre for Theoretical Physics,
Jamia Millia Islamia, New Delhi 110025, India}
\affiliation{Astrophysics and Cosmology Research Unit, School of Mathematics, Statistics and Computer Science,\\
University of KwaZulu-Natal, Private Bag 54001, Durban 4000, South Africa}
\author[0000-0001-5545-3507]{Misba Afrin}
\affiliation{Centre for Theoretical Physics,
Jamia Millia Islamia, New Delhi 110025, India}
\correspondingauthor{Misba Afrin}
\email{me.misba@gmail.com}
\begin{abstract}
The Event Horizon Telescope (EHT) recently released an image of the supermassive black hole Sgr A* showing an angular shadow diameter $d_{sh}= 48.7 \pm 7\,\mu$as and Schwarzschild shadow deviation $\delta = -0.08^{+0.09}_{-0.09}~\text{(VLTI)},-0.04^{+0.09}_{-0.10}~\text{(Keck)}$ using a black hole mass $M = 4.0^{+1.1}_{-0.6} \times 10^6 M_\odot $. The EHT image of Sgr A* is consistent with a Kerr black hole's expected appearance, and the results directly prove the existence of a supermassive black hole at the center of the Milky Way. Here, we use the EHT observational results for Sgr A* to investigate the constraints on its charge with the aid of Kerr-like black holes, paying attention to three leading rotating models, namely Kerr--Newman, Horndeski, and hairy black holes. Modeling the supermassive black hole Sgr A* as these Kerr-like black holes, we observe that the EHT results of Sgr A* place more strict upper limits on the parameter space of Kerr--Newman and Horndeski black holes than those placed by the EHT results for M87*. A systematic bias analysis reveals that, observational results of future EHT experiments place more precise limits on the charge of black hole Sgr A*. Thus, the Kerr-like black holes and Kerr black holes are indiscernible in a substantial region of the EHT-constrained parameter space; the claim is substantiated by our bias analysis.
\end{abstract}

\section{Introduction}
According to the Kerr hypothesis, the Kerr metric (\citeyear{Kerr:1963ud}) is the only stationary, vacuum, axisymmetric solution of Einstein's field equations that does not have pathologies outside the event horizon and is asymptotically flat \citep{Israel:1967wq,Carter:1971zc,Hawking:1971vc}. But direct evidence for the theorem is still inconclusive, and it may be difficult to rule out other Kerr-like black holes \citep{Ryan:1995wh,Will:2005va} that are admitted by modified theories of gravity. Also, there remain, inconclusive fundamental issues in general relativity (GR), at scales comparable to the event horizon.

The event horizon is accessible only for indirect tests with strong-field phenomena \citep{Falcke:1999pj}, such as the black hole shadow \citep{Bardeen:1973tla,Luminet:1979nyg}, and one of the first quantitative suggestions for performing tests of the Kerr metric with the black hole shadows was given by \cite{Johannsen:2010ru}. Black holes became a physical reality in 2019 with the release of the first horizon-scale image of the black hole M87* by the Event Horizon Telescope (EHT) collaboration \citep{EventHorizonTelescope:2019dse,EventHorizonTelescope:2019ggy,EventHorizonTelescope:2019jan,EventHorizonTelescope:2019pgp,EventHorizonTelescope:2019ths,EventHorizonTelescope:2019uob}, and bounds could be placed on the size of the compact emission region size with angular diameter $\theta_d=42\pm 3\, \mu $as along with the central flux depression with a factor of $\gtrsim 10$, which can be identified as the shadow \citep{EventHorizonTelescope:2019dse,EventHorizonTelescope:2019pgp,EventHorizonTelescope:2019ggy}. 
In turn, the EHT collaboration, in 2022, released the image of the black hole Sgr A*  in the Milky Way showing an angular shadow of diameter $d_{sh}= 48.7 \pm 7\,\mu$as, considering stellar dynamical priors on its mass and distance \citep{EventHorizonTelescope:2022exc,EventHorizonTelescope:2022urf,EventHorizonTelescope:2022vjs,EventHorizonTelescope:2022wok,EventHorizonTelescope:2022xnr,EventHorizonTelescope:2022xqj}.
The observed images of the two black holes M87* and Sgr A* are consistent with the expected appearance of a Kerr black hole as predicted by GR \citep{EventHorizonTelescope:2019dse,EventHorizonTelescope:2019ggy,EventHorizonTelescope:2022xnr}.  Nevertheless, the current uncertainty in the measurement of spin or angular momentum and the relative deviation of quadrupole moments do not eliminate Kerr-like black holes arising in modified gravities \citep{EventHorizonTelescope:2019dse,EventHorizonTelescope:2019pgp,EventHorizonTelescope:2019ggy,Cardoso:2019rvt}.
Furthermore, when compared with the EHT results for  M87*, the Sgr A* exhibits consistency with the predictions of GR stretching across three orders of magnitude in central mass \citep{EventHorizonTelescope:2022xnr}.

The EHT bounds on the observables of the black hole shadow of Sgr A* provide an excellent way to examine the viability of various non-Kerr black hole models, with additional deviation parameters or charges to Kerr black holes, and to constrain the charges \citep{Khodadi:2022pqh,Vagnozzi:2022moj,KumarWalia:2022ddq,KumarWalia:2022aop,Uniyal:2022vdu}. While the EHT measurements contain far more information related to the image of Sgr A*, for our purpose, we shall consider only the bounds on the shadow observables, i.e., the angular shadow diameter $d_{sh}$ and Schwarzschild shadow deviation $\delta$ \citep{EventHorizonTelescope:2019dse}, to put constraints on the \emph{charges} (or \emph{hairs}) of the three well known rotating black hole metrics. We also investigate whether the bounds for Sgr A* can provide more stringent constraints on the black hole charges than previously obtained with the bounds for M87* observables \citep{EventHorizonTelescope:2020qrl,Afrin:2021imp,Ghosh:2020spb,EventHorizonTelescope:2021dqv,Afrin:2021wlj,Kumar:2020owy}. Within the EHT-constrained parameter space, we conduct a systematic bias analysis between the Kerr-like black hole shadows taken as models and Kerr shadows as injection, to determine how the limits on charges placed with the current EHT observations of Sgr A* will be affected, as and when future more precise observational data are obtained \citep{EventHorizonTelescope:2022xqj}.
\begin{figure*}[t]
\centering
\begin{tabular}{c c c}
    \includegraphics[scale=0.53]{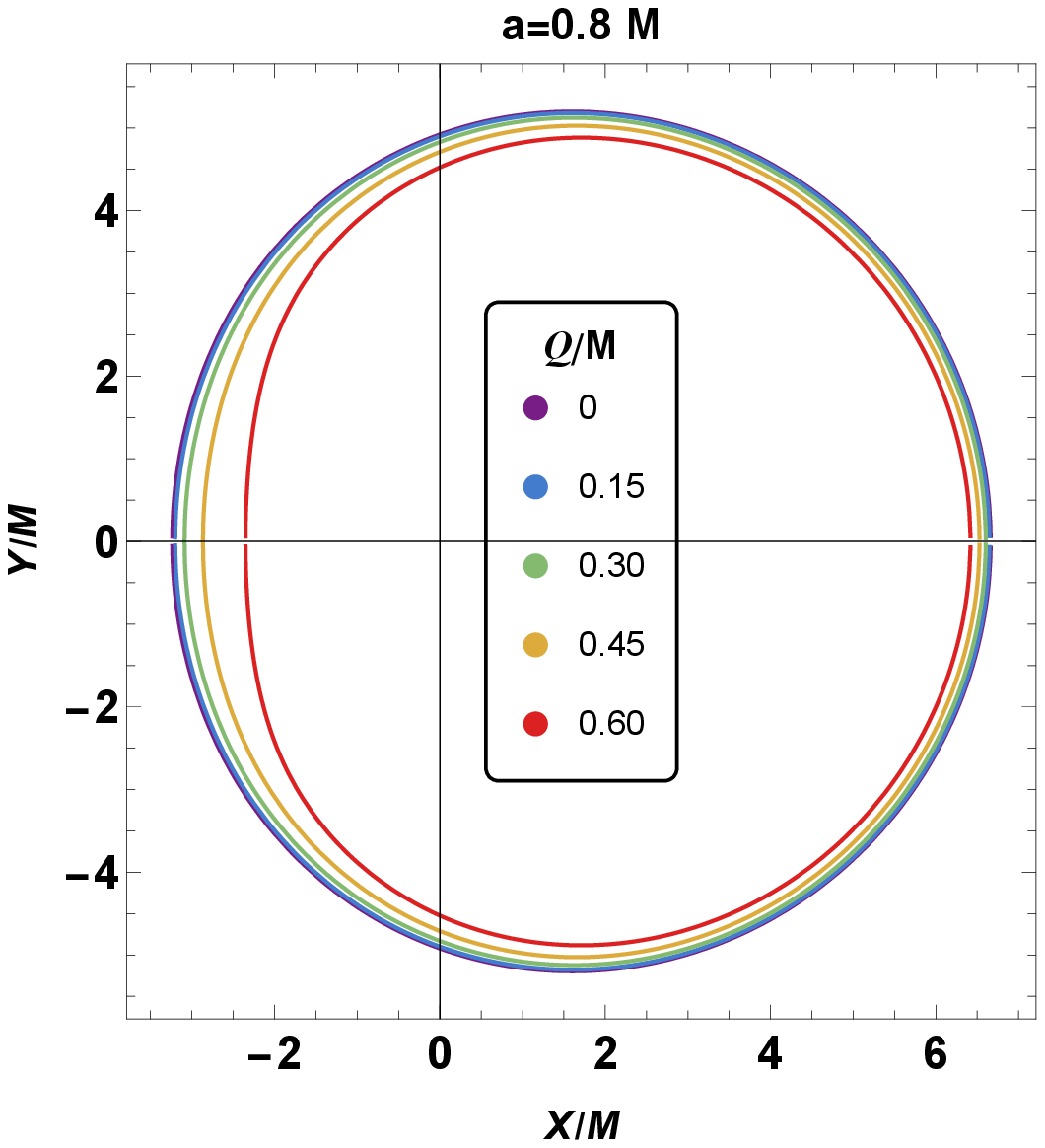}&
    \includegraphics[scale=0.53]{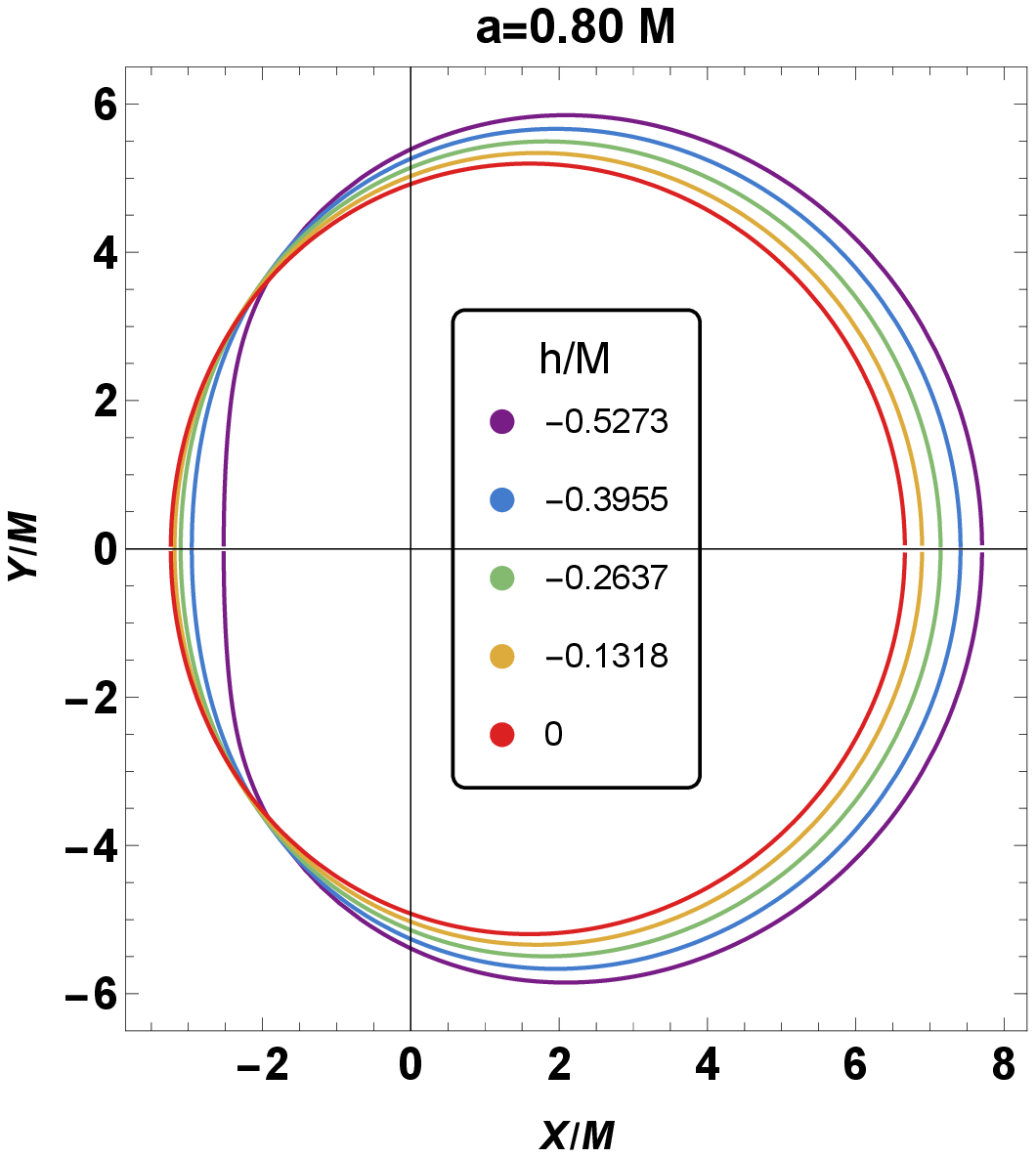}&
    \includegraphics[scale=0.53]{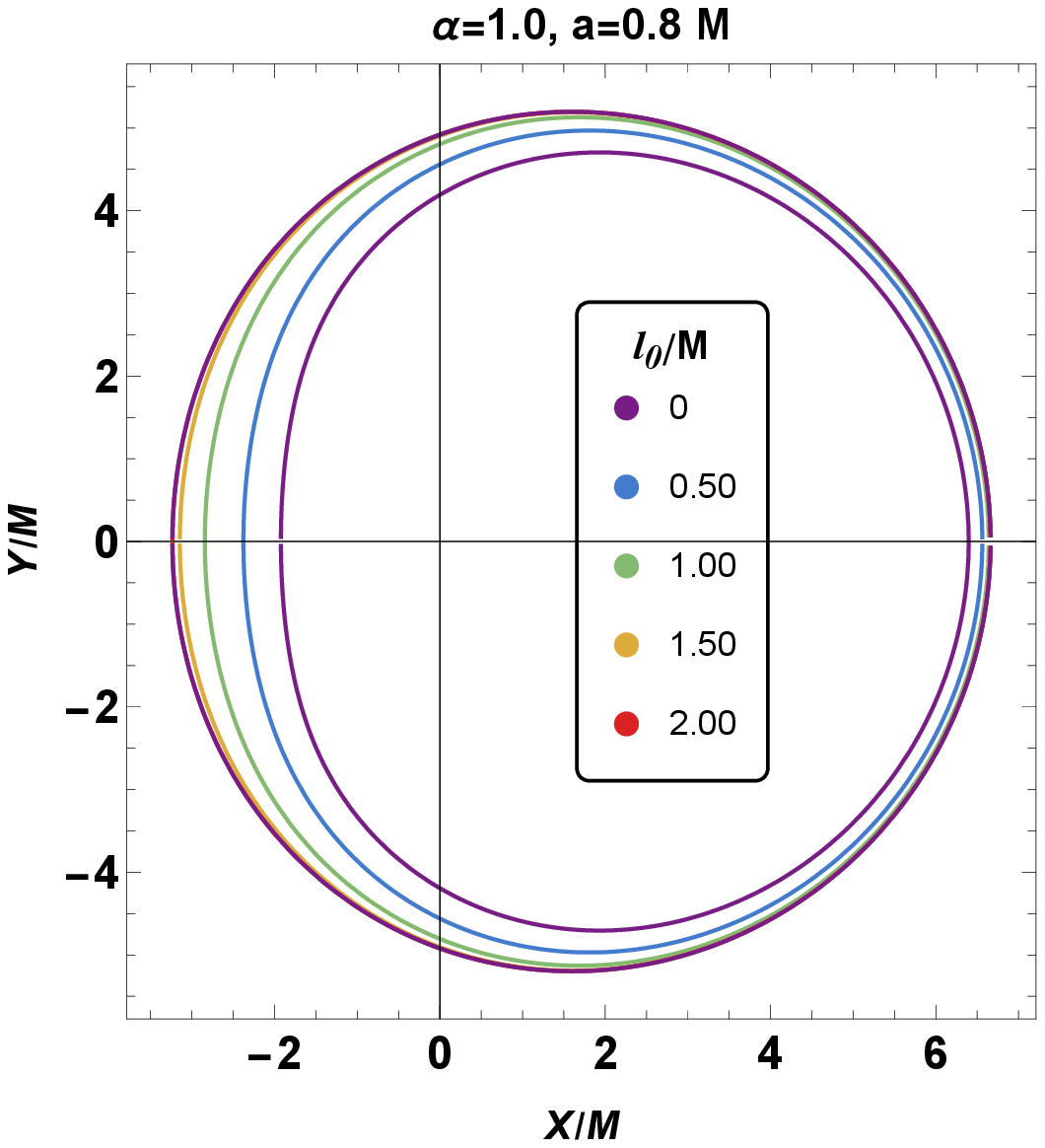}
\end{tabular}
\caption{Shadows of Kerr-like black holes: Kerr--Newman (left), rotating Horndeski \citep[middle,][]{Afrin:2021wlj} and rotating hairy \citep[right,][]{Afrin:2021imp} black holes with varying charges. The solid red curves correspond to Kerr black hole shadows.}\label{shadow_Figure}
\end{figure*}

We work with geometrized units $8 \pi G=c=1$ throughout this paper, unless units are specifically defined.

\section{Shadow of Kerr-like black holes}\label{sect2}
We begin by exploring analytically the shadow features of the  Kerr-like metric---an asymptotically flat,
stationary, and axisymmetric spacetime, whose line element in Boyer--Lindquist coordinates reads \citep{Bambi:2013ufa,Tsukamoto:2017fxq,Kumar:2020yem}
\begin{eqnarray}\label{metric}
ds^2 & = & - \left( 1- \frac{2m(r)r}{\Sigma} \right) dt^2  - \frac{4am(r)r}{\Sigma  } \sin^2 \theta\, dt \, d\phi +
\frac{\Sigma}{\Delta}dr^2  \nonumber
\\ && + \Sigma\, d \theta^2+ \left[r^2+ a^2 +
\frac{2m(r) r a^2 }{\Sigma} \sin^2 \theta
\right] \sin^2 \theta\, d\phi^2,
\end{eqnarray}
and
\begin{equation}
\Sigma = r^2 + a^2 \cos^2\theta,\;\;\;\;\;  \Delta = r^2 + a^2 - 2m(r)r,\label{Delta}
\end{equation}
where $m(r)$ is the mass function such that $\lim_{r\to\infty}m(r)=M$ with $M$ being the ADM mass of the rotating black hole and $a$ is the spin. We assume that the $m(r)$ is well behaved for $r>r_+$, where $r_+$ is the radius of the event horizon. The metric (\ref{metric}), in general, depending on the choice of mass function $m(r)$, describes a wide variety of rotating black holes \citep{Tsukamoto:2017fxq,Kumar:2018ple,Afrin:2021imp,Afrin:2021wlj}.

For an asymptotically distant observer ($r_0\to\infty$), making an inclination angle $\theta_0$ with the spin axis, the black hole shadow is a dark region in the celestial sky outlined by a bright ring \citep{Johannsen:2015mdd,Johnson:2019ljv} with Cartesian coordinates \citep{Bardeen:1973tla,Afrin:2021wlj}
\begin{equation}
\{X,Y\}=\{-\xi_{c}\csc\theta_0,\, \pm\sqrt{\eta_{c}+a^2\cos^2\theta_0-\xi_{c}^2\cot^2\theta_0}\}\,.\label{pt}
\end{equation} 
where 
where $\xi_c$ and $\eta_c$ are the critical impact parameters of the unstable spherical photon orbits hitting the observer's celestial plane and can be determined as a function of $m(r)$ and $m'(r)$ \citep{Tsukamoto:2017fxq,Kumar:2020yem,Kumar:2020ltt}.
 \begin{figure*}[t]
\begin{center}
    \begin{tabular}{c c}
    \hspace{-1cm}\includegraphics[scale=0.80]{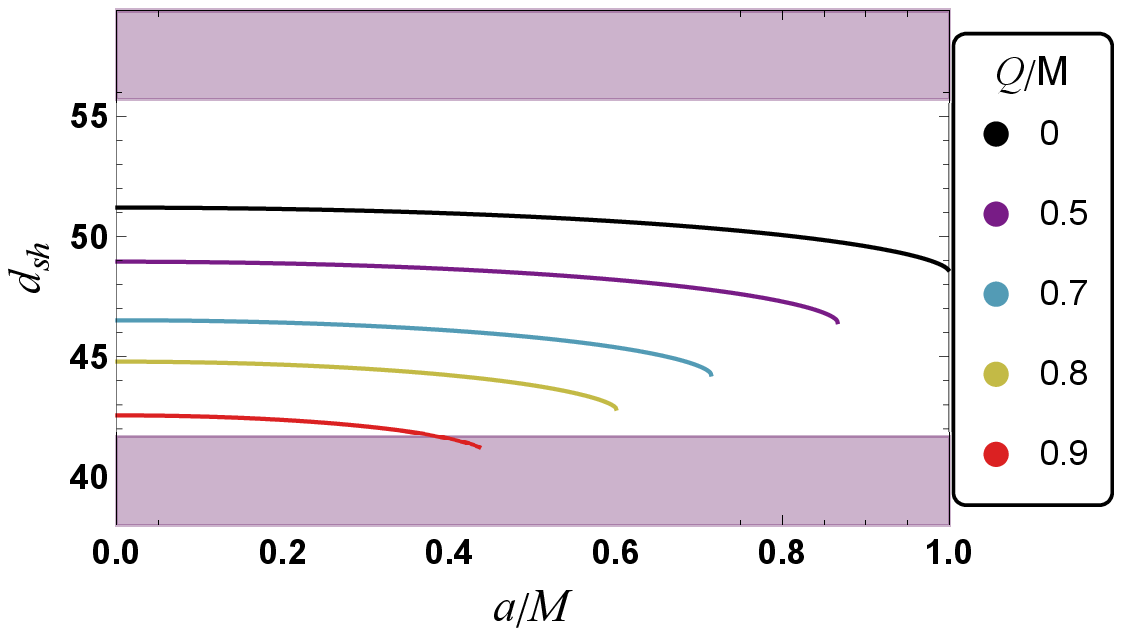}&
     \hspace{-0.3cm}\includegraphics[scale=0.80]{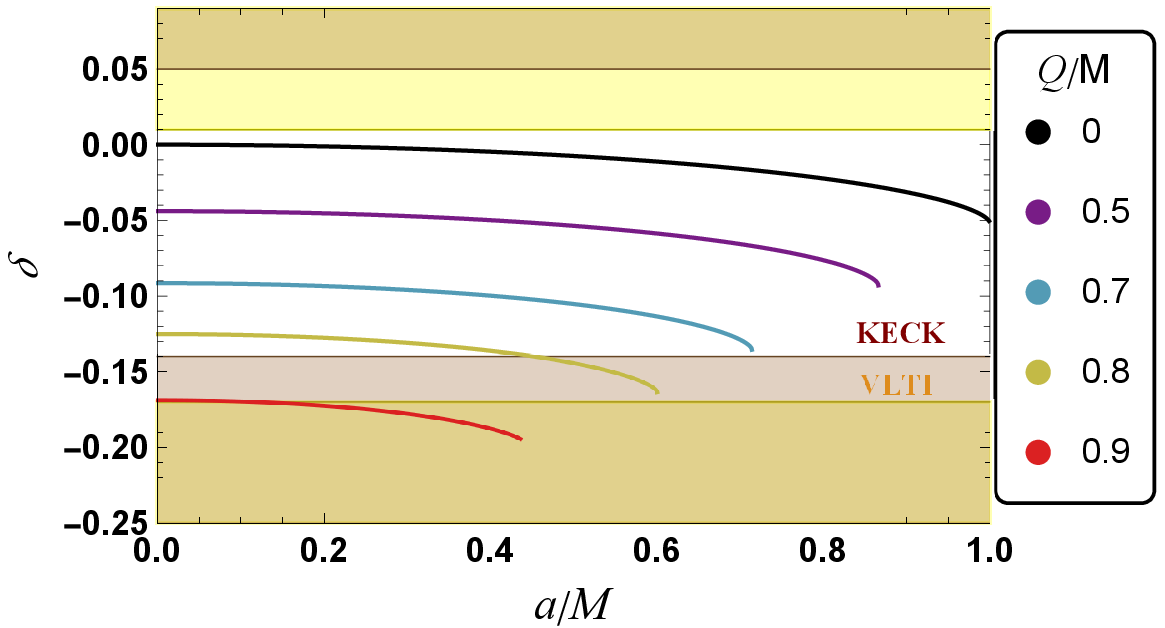}\\
     \hspace{-1cm}\includegraphics[scale=0.80]{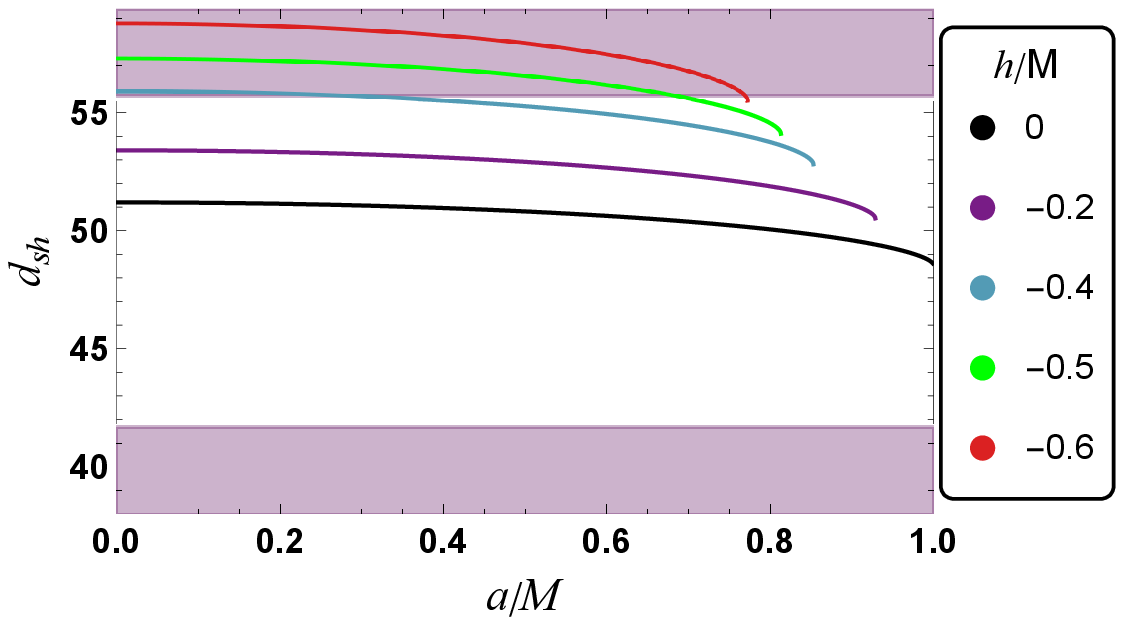}&
     \hspace{-0.3cm}\includegraphics[scale=0.80]{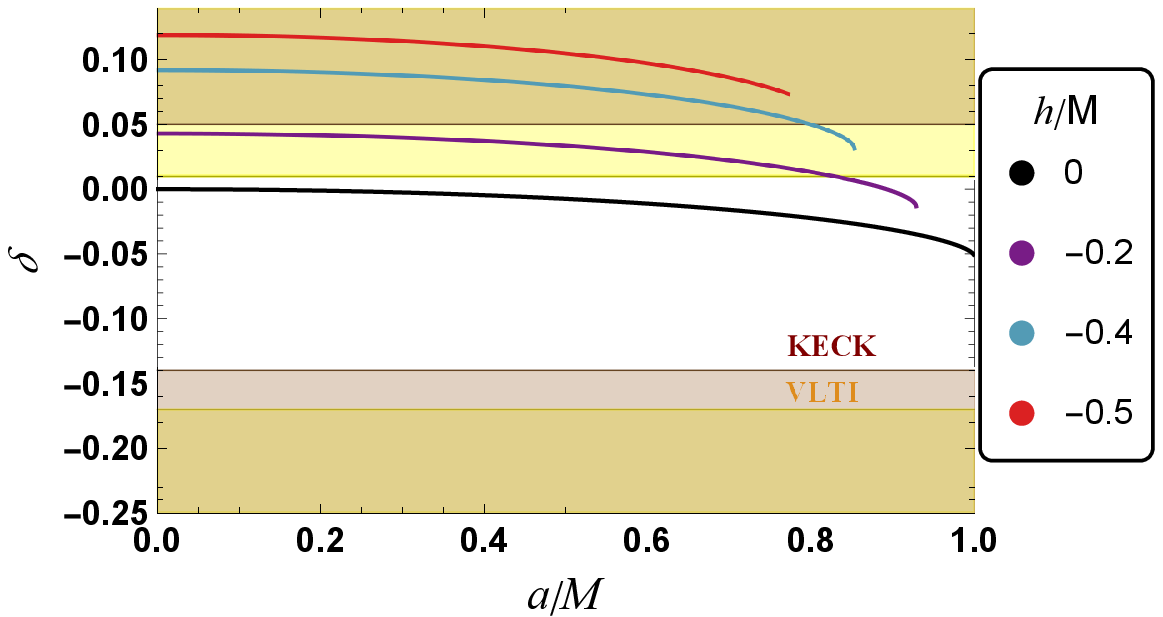}\\
     \hspace{-1cm}\includegraphics[scale=0.80]{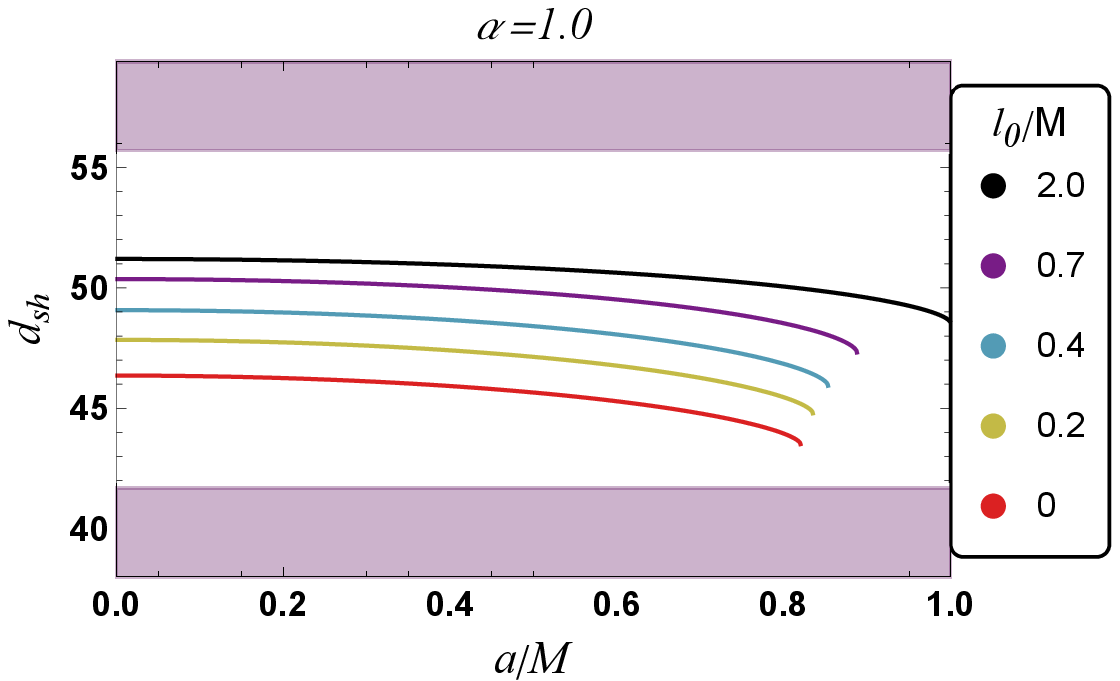}&
     \hspace{-0.3cm}\includegraphics[scale=0.80]{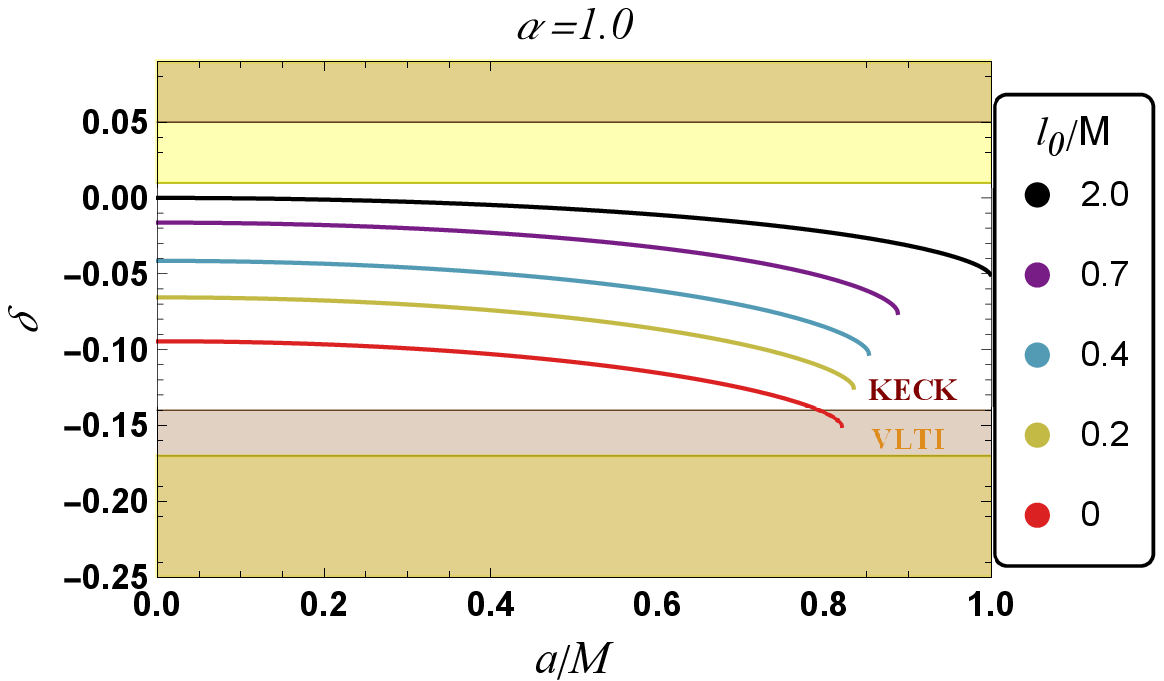}
\end{tabular}
\end{center}
	\caption{Behaviour of shadow angular diameter (left) and Schwarzschild shadow deviation (right) of Kerr-like black shadows, viz. Kerr--Newman black hole (\textit{top}),  rotating Horndeski black hole (middle) and rotating hairy  black hole (\textit{bottom}). In the white regions, the shadow observables are consistent within $68\%$ confidence level of the EHT results of Sgr A*. The shadow observables have a weaker dependence on spin $a$ than on charge.}
	\label{ObservablesPlot}
\end{figure*}
 \begin{figure*}[t]
\begin{center}
    \begin{tabular}{c c}
    \includegraphics[scale=0.80]{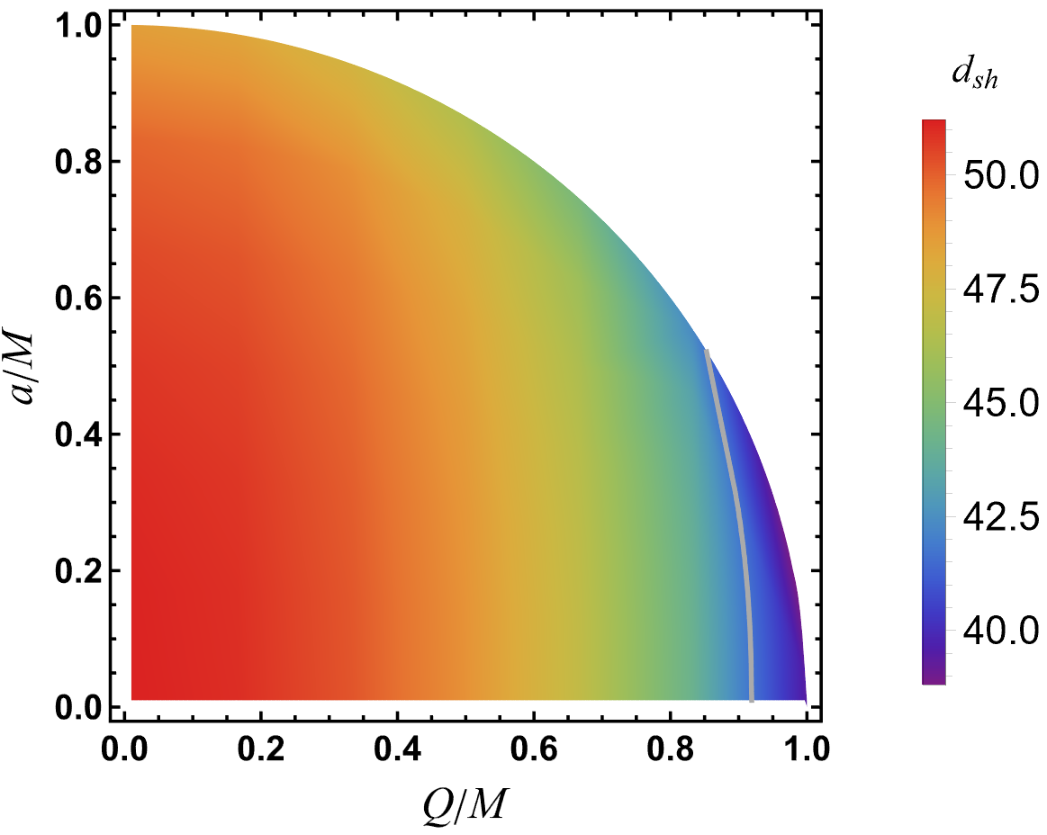}&
     \includegraphics[scale=0.80]{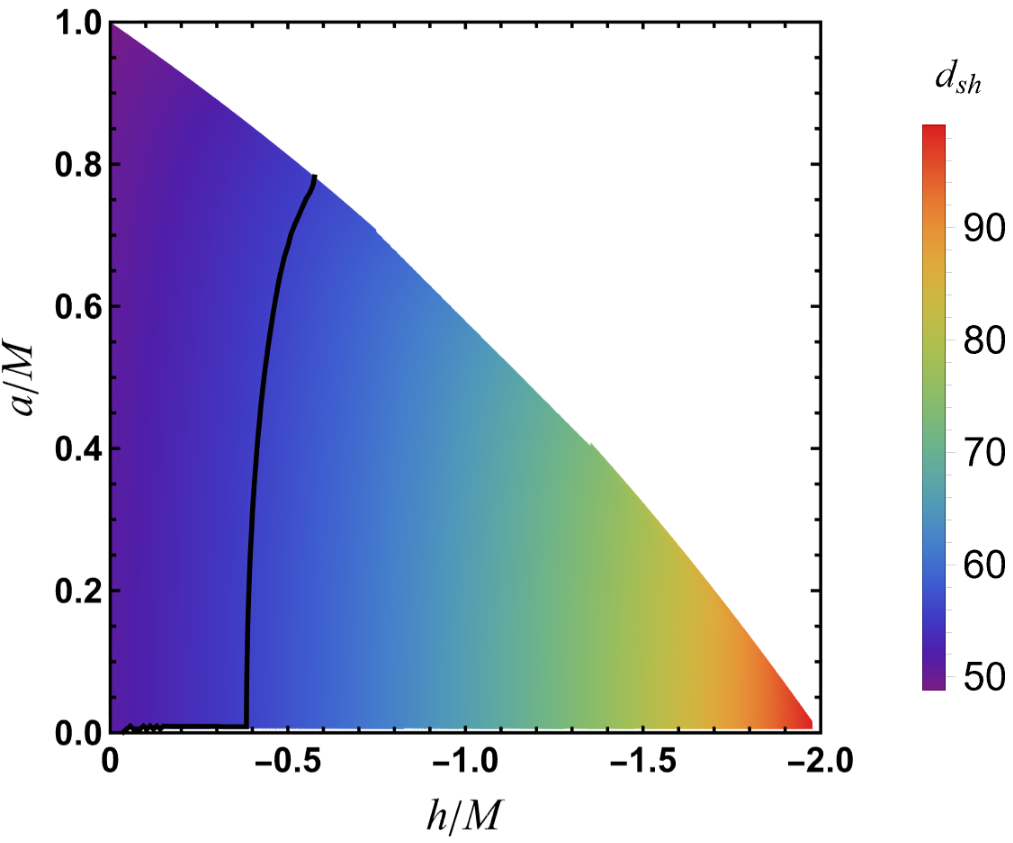}\\
     \includegraphics[scale=0.80]{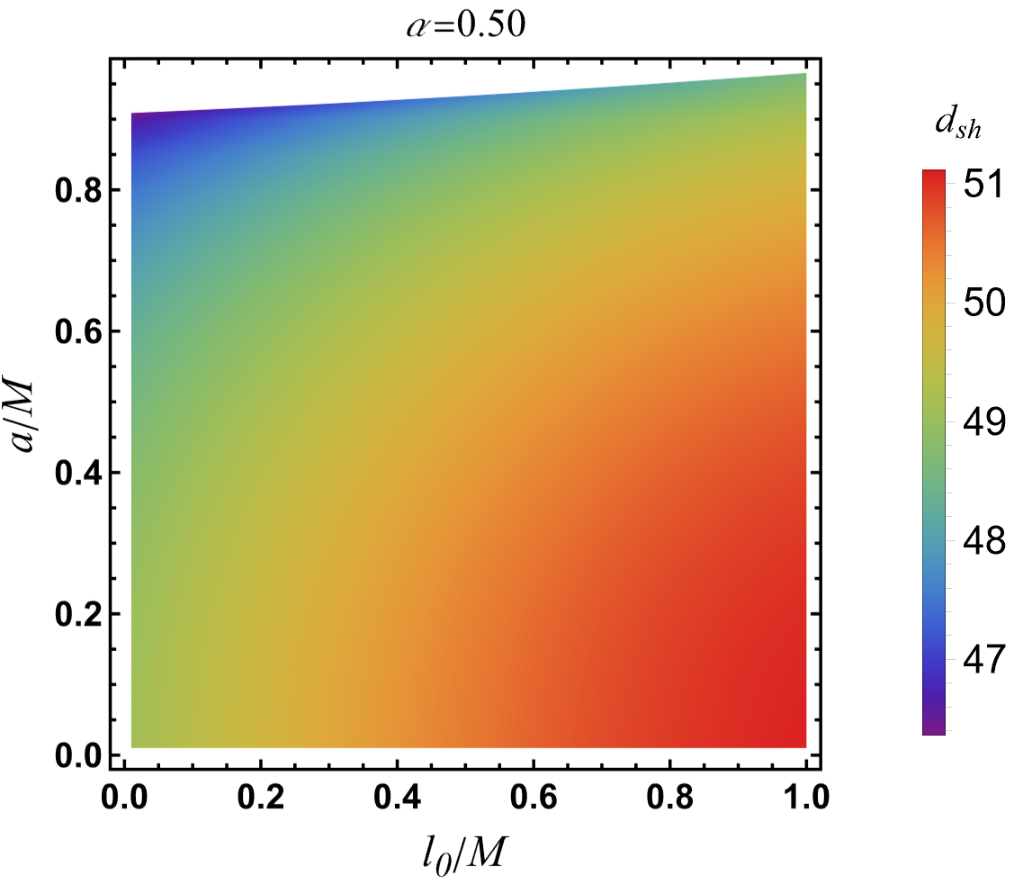}&
     \includegraphics[scale=0.80]{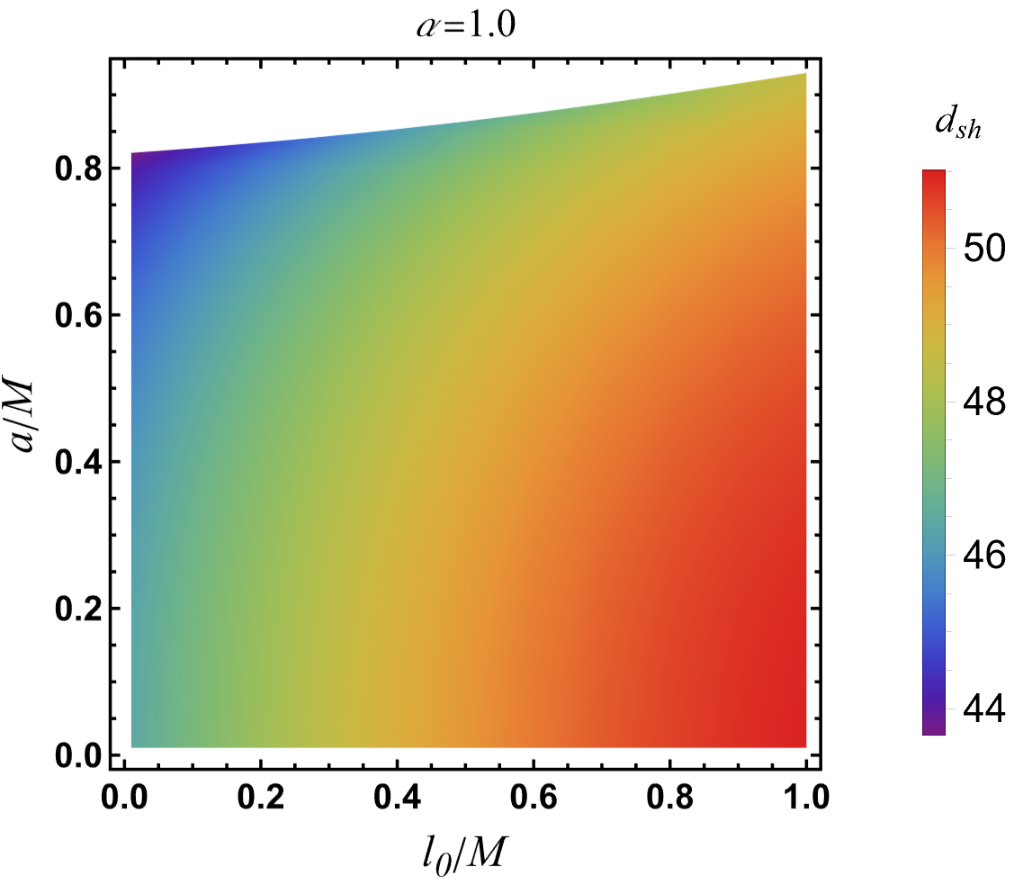}
\end{tabular}
\end{center}
	\caption{Constraints from EHT results of angular shadow diameter  $d_{sh}$: Modeling Sgr A* as Kerr-like black holes, viz.  Kerr--Newman  (\textit{top left}),  rotating Horndeski (\textit{top right}) and rotating hairy (\textit{bottom}) black holes. The grey and black curves correspond respectively to $41.7\mu$as and $55.7\mu$as, and the parameter space bounded within these curves and the axes is consistent with the current EHT results of Sgr A*. The entire parameter space for the rotating hairy black hole is consistent with the shadow angular diameter of Sgr A*. The white region corresponds to the forbidden parameter space.}
	\label{AngDiamFigure}
\end{figure*}
We examine here three Kerr-like black holes, which are defined by the metric (\ref{metric}) with appropriate mass function $m(r)$.
\paragraph{Kerr--Newman black holes.}
The mass function of the Kerr--Newman metric is given by
\begin{equation}\label{KN_mass}
    m(r)= M-\frac{Q^2}{2r},
\end{equation}
where $Q$ is the electric charge of the black hole.
Though astrophysical black holes are supposed to be neutral  \citep{Gibbons:1975kk}, there is a still good reason to constrain the electric charge of black holes  \citep{Takahashi:2005hy,EventHorizonTelescope:2021dqv,EventHorizonTelescope:2022xqj}. Further, a small nonzero charge may be accumulated by rotating black holes \citep{Zajacek:2019kla} alongside the induction of electric charge \citep{Wald:1974np,deDiego:2004ar,Levin:2018mzg}. Besides, the charge neutrality of astrophysical black holes has not been yet confirmed observationally. The EHT collaboration has considered Reissner–-Nordstr\"{o}m and Kerr--Newman black holes to put constraints on the electric charge, besides other charged black holes considered previously \citep{EventHorizonTelescope:2021dqv}. Modeling M87* as Kerr--Newman black hole puts constraints on the parameter $Q/M\in (0,0.90]$ \citep{EventHorizonTelescope:2021dqv}.
The charge of Sgr A* has also been theoretically constrained to be $Q\leq 3.1 \times 10^8$C using Chandra X-ray data \citep{Karouzos2018,Zajacek:2018ycb}. We intend here to check whether better constraints can be placed on the charge of Sgr A* with the recent EHT results, and thus aim to establish or discard the charge neutrality of the astrophysical black holes, observationally, at the current and future resolutions of the EHT.
\paragraph{Rotating Horndeski black holes.}
The Horndeski black holes are an exact solution to a class of quartic Horndeski gravity that is asymptotically flat \citep{Bergliaffa:2021diw,Kumar:2021cyl}. Astrophysical probes of these black holes draw interest due to the fact that the Horndeski theory, a ghost-free scalar--tensor theory, gives an alternative explanation for dark energy \citep{Kase:2018aps}. The rotating Horndeski black hole \citep{Afrin:2021wlj,Kumar:2021cyl} is described by the metric (\ref{metric}) with the mass function 
\begin{equation}\label{Horndeski_mass}
    m(r)=M-\frac{h}{2}\ln\left(\frac{r}{2M}\right),
\end{equation}
where the deviation charge parameter $h$ comes from the Horndeski theory \citep{Bergliaffa:2021diw}.  
Modeling M87* as rotating Horndeski black holes puts constraints on the parameters, namely $0.0077 M\leq a\leq 0.9353 M$, $-0.7564 M\lesssim h<0$ at $\theta_o=90$\textdegree\, and $0.0048 M\leq a\leq 0.9090 M$, $-0.7920 M\lesssim h<0$ at $\theta_o=17$\textdegree \citep{Afrin:2021wlj}.

\paragraph{Rotating hairy black holes.}
The rotating hairy black hole, generated by the gravitational decoupling method \citep{Contreras:2021yxe}, is a modified Kerr black hole solution with a surrounding fluid, which can be any viable form of matter--energy distribution, including dark matter, and is described by (\ref{metric}) with mass function \citep{Contreras:2021yxe}
\begin{equation}\label{HairyKerr_mass}
    m(r)=M-\frac{\alpha}{2}r e^{-r/(M-\frac{l_0}{2})},
\end{equation}
where the parameter $l_0=\alpha l\leq2M$ is the charge, which determines the asymptotic flatness of metric (\ref{metric}) with (\ref{HairyKerr_mass}). $l_0$ has been called a primary hair \citep{Contreras:2021yxe}. Also, in both the limits $\alpha=0$ and $l_0 \to  l_k=2M$ one recovers the Kerr black hole. Modeling  rotating hairy black holes as M87* yields constraints on the parameters $l_0/M \in[0.7122,1)$ \citep{Afrin:2021imp}. 

\begin{figure*}[t]
\begin{center}
    \begin{tabular}{c c}
    \includegraphics[scale=0.80]{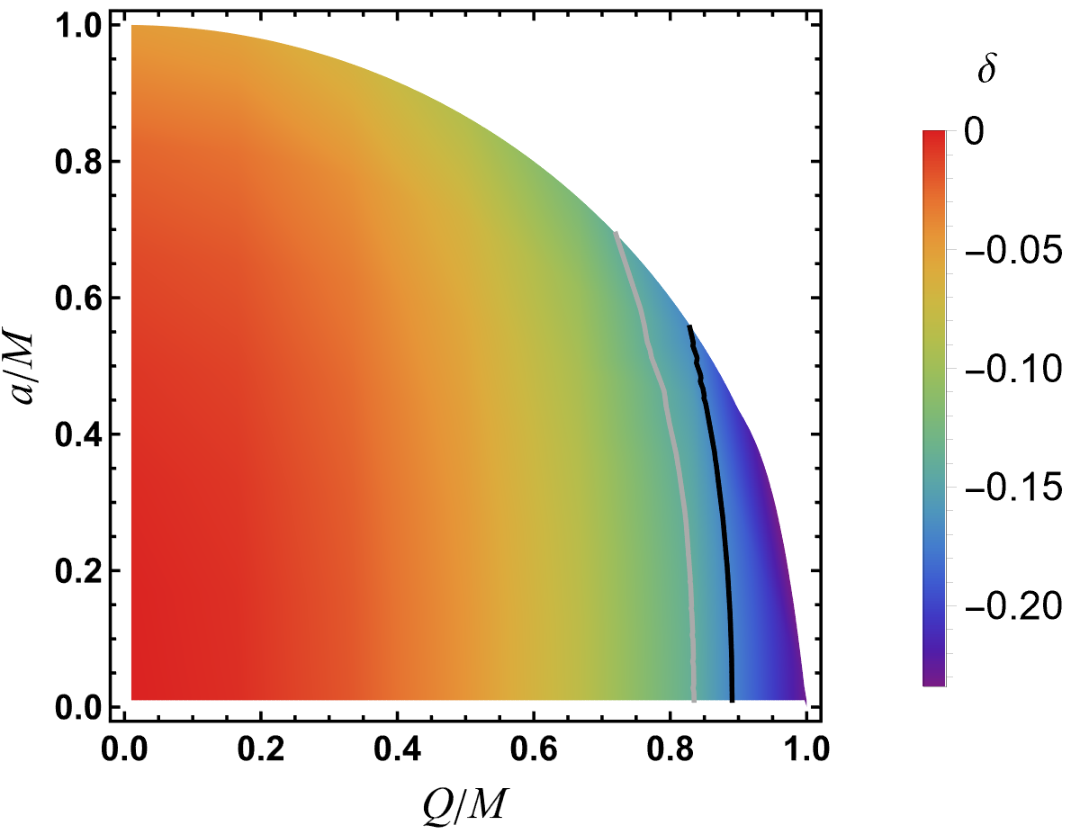}&
     \includegraphics[scale=0.80]{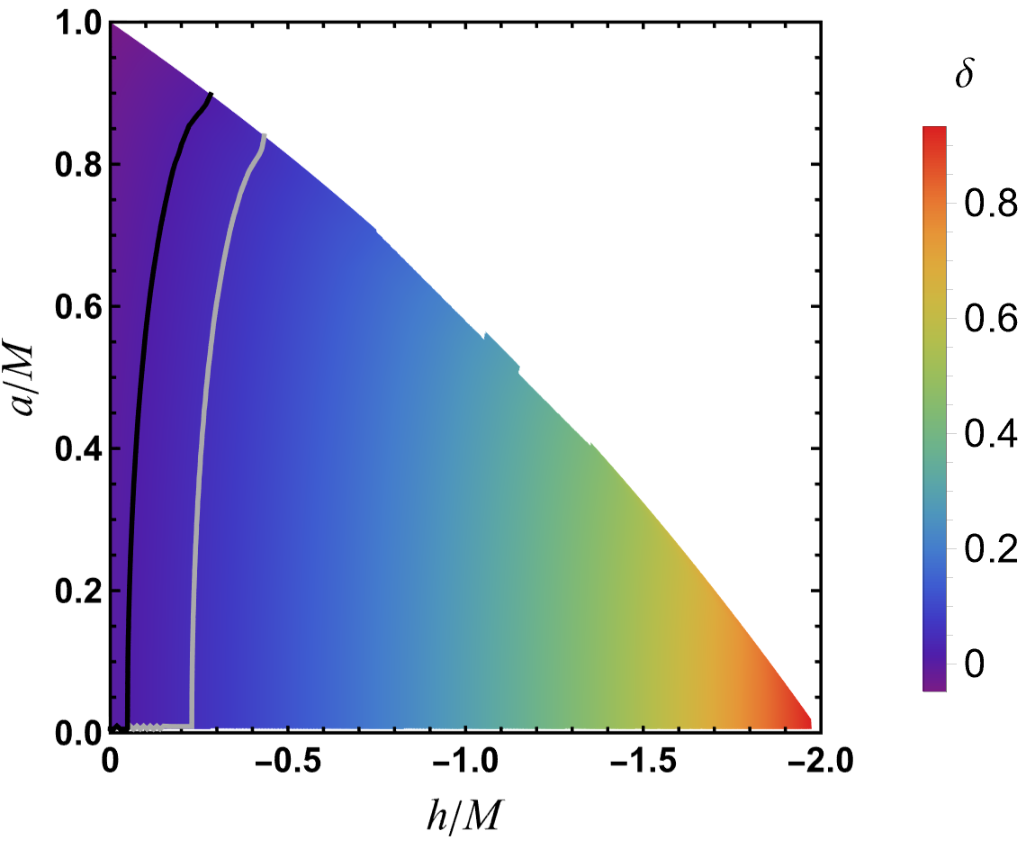}\\
     \includegraphics[scale=0.80]{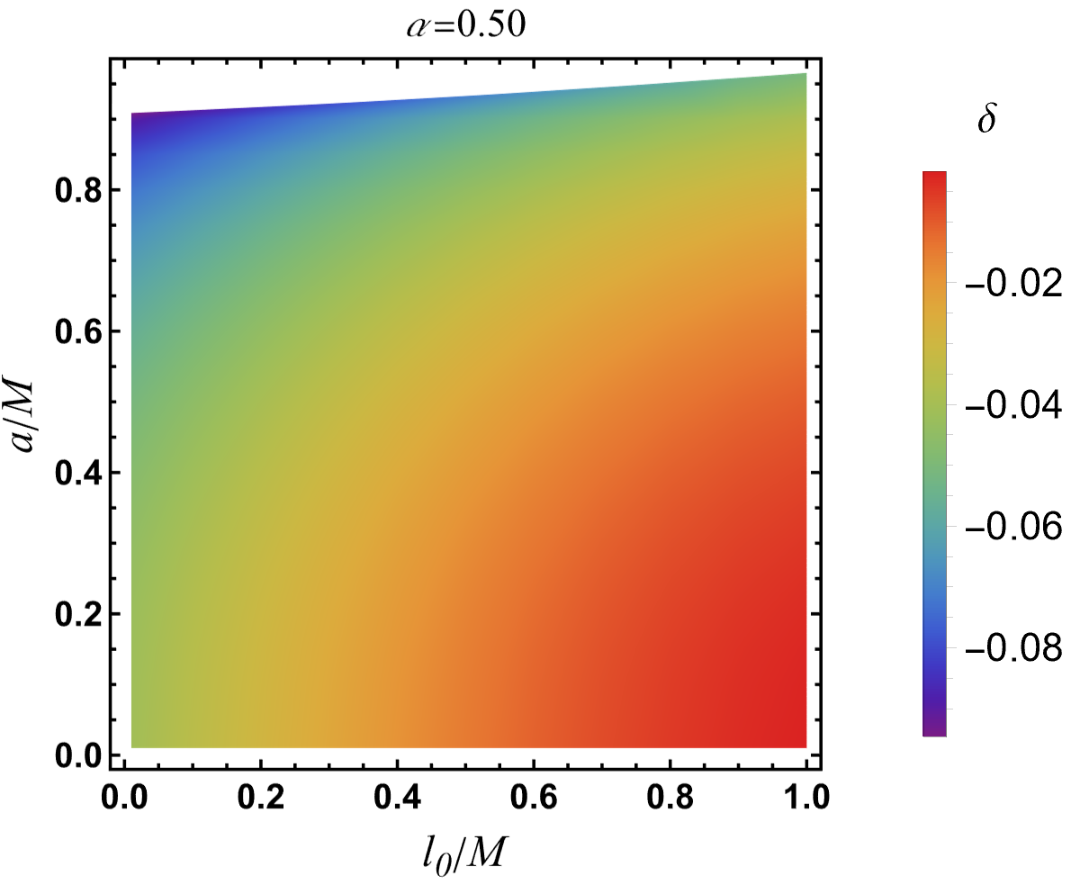}&
     \includegraphics[scale=0.80]{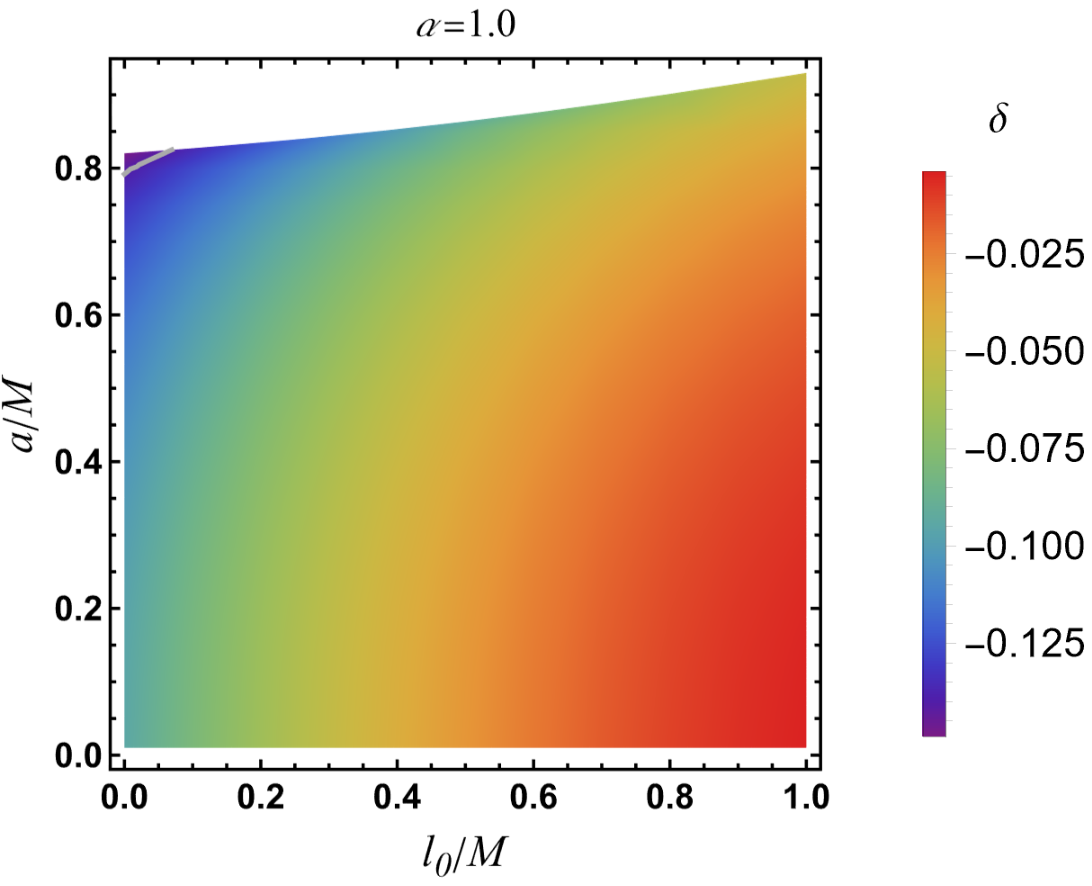}
\end{tabular}
\end{center}
	\caption{Constraints from EHT results of Schwarzschild shadow deviation $\delta$:  Modeling Sgr A* as Kerr--Newman black hole (\textit{top left}),  rotating Horndeski black hole (\textit{top right}) and rotating hairy  black hole (\textit{bottom}). The gray and black curves correspond respectively to Keck and VLTI bounds, and the parameter space bounded within these curves and the axes is consistent with the current EHT results of Sgr A*. The entire parameter space for the rotating hairy black hole is consistent with Schwarzschild shadow deviation of Sgr A* for $\alpha=0.50$. The white region corresponds to forbidden parameter space.}
	\label{ShadowDeviationFigure}
\end{figure*}
\begin{figure}[t]
    \centering
    \includegraphics[scale=0.8]{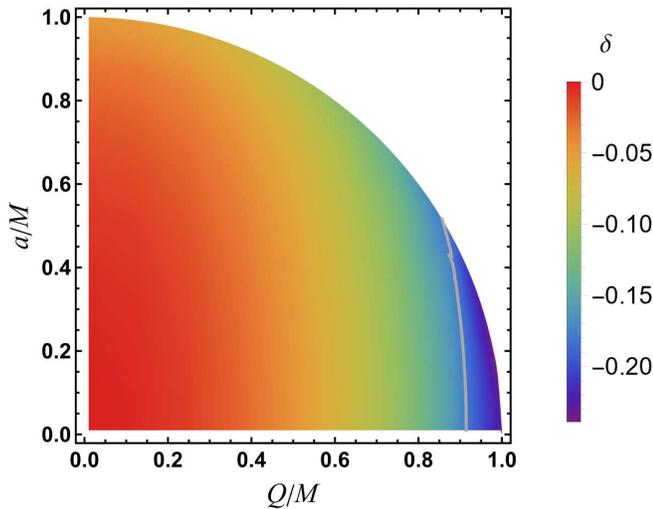}
    \caption{Constraints from EHT results of Schwarzschild shadow deviation Modeling M87* as Kerr--Newman black hole. The gray curve correspond to $\delta=-0.18$, and the parameter space bounded within this curve and the axes is consistent with the current EHT results of M87*. The white region corresponds to forbidden parameter space.}
    \label{KN_M87}
\end{figure}
\begin{figure*}[t]
\begin{center}
    \begin{tabular}{c c}
   \hspace{-0.6cm} \includegraphics[scale=0.90]{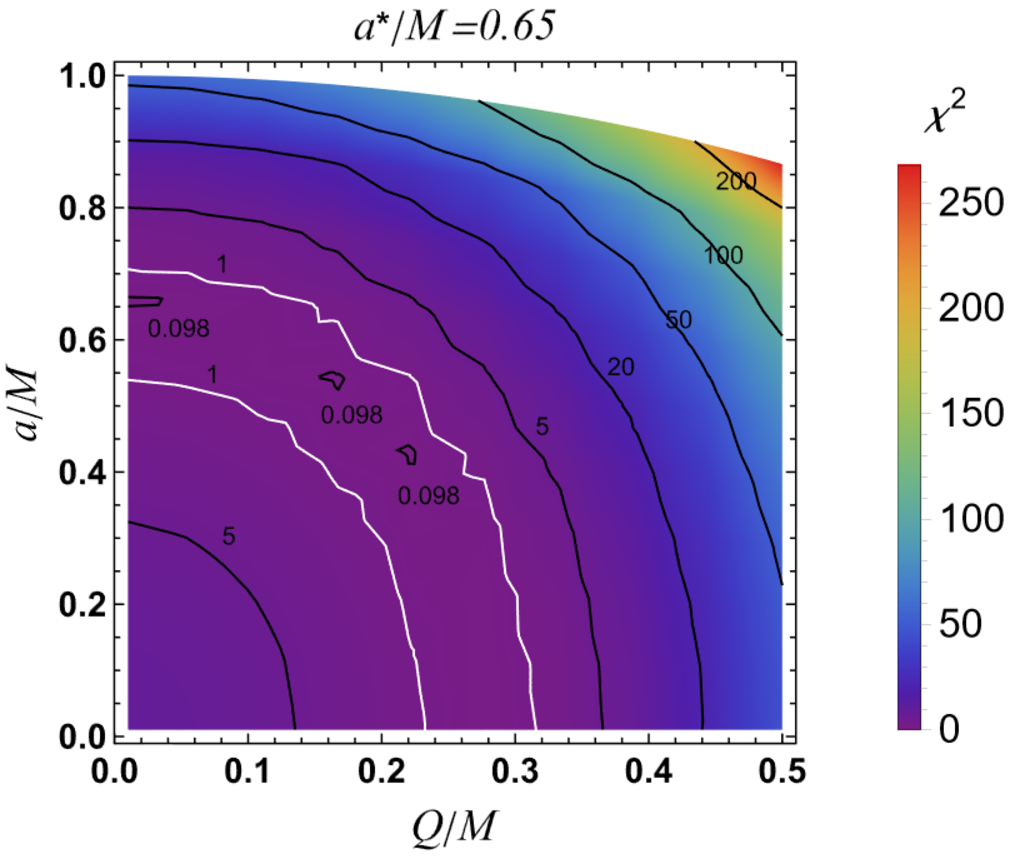}&
     \includegraphics[scale=0.90]{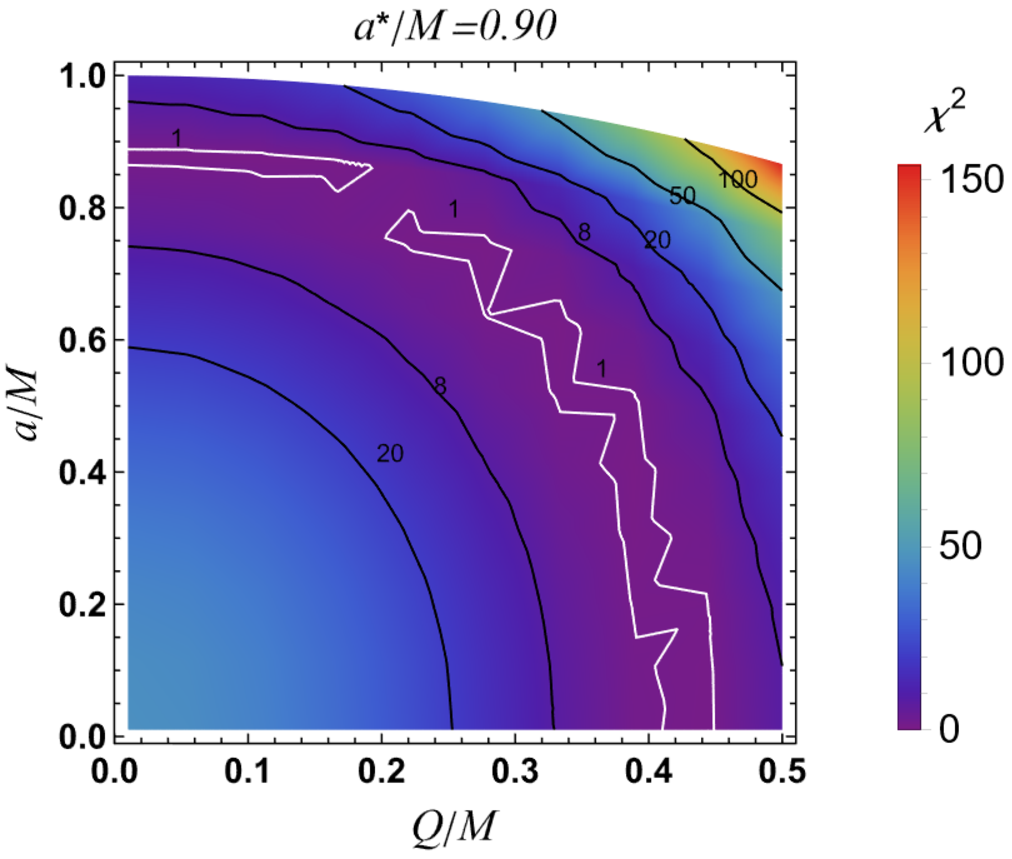}
\end{tabular}
\end{center}
	\caption{Reduced $\chi^2$ between injected Kerr shadow and modeled Kerr--Newman black hole shadow in the parameter space constrained with EHT results for Sgr A* for injected spin $a^*=0.65 M$ (left) and $0.90 M$ (right). The white line denotes the $\chi^2=1$ contour and  the white region is a forbidden parameter space.}
	\label{ChiSqrFigure1}
\end{figure*}
\section{Constraints from EHT results on Sgr A*}\label{sect3}
The shape and size of the black hole shadow serve as a direct probe of strong-field gravity, as it is the most direct manifestation of the background spacetime, notwithstanding multifarious astrophysical processes such as accretion flow, emission phenomena, etc. \citep{Afrin:2021wlj}. This section presents constraints on the charge parameters associated with the three Kerr-like black holes, imposed by the recently released images of Sgr A* recently released by the EHT collaboration \citep{EventHorizonTelescope:2022xnr,EventHorizonTelescope:2022xnr} on the observables relating to the characteristic shadow size.
\begin{table*}[t]
\caption{Constraints on Charges from EHT Observation of Sgr A*}
      \centering
\begin{tabular}{|l|ll|}
\hline
\multirow{2}{*}{Spacetimes} & \multicolumn{2}{l|}{\quad\quad\quad\quad\quad Constrains from EHT Observations } \\ \cline{2-3} 
                        & \multicolumn{1}{l|}{\quad\quad\quad$d_{sh}$}    & \quad\quad\quad\quad\quad\quad$\delta$    \\ \hline
Kerr--Newman & \multicolumn{1}{l|}{$Q/M\in (0, 0.8478]$}  & \begin{tabular}[c]{@{}l@{}}\hspace{-1cm}VLTI: $Q/M\in (0, 0.8255]$\\ \hspace{-1cm}Keck: $Q/M\in (0, 0.7174]$\end{tabular}    \\ \hline
Rotating Horndeski   & \multicolumn{1}{l|}{$h/M\in [-0.3804,0)$} & \begin{tabular}[c]{@{}l@{}}\hspace{-1cm}VLTI: $h/M\in [-0.0475, 0)$    \\ \hspace{-1cm}Keck: $h/M\in [-0.2325, 0)$\end{tabular} \\ \hline
Rotating hairy               & \multicolumn{1}{l|}{$l_0/M\in [0, 1)$}         & {\;Keck: $l_0/M\in [0.0696,1)$}         \\ \hline
\end{tabular}
\label{parameters_Table}
\end{table*}

Previously, the EHT used an extensive library of ray-traced general-relativistic magnetohydrodynamic simulations of black holes, had derive a central mass $M_{M87^*}=(6.5\pm0.7)\times 10^9 M_\odot$ and distance $d_{M87^*}=16.8$ Mpc. The mass and distance of Sgr A* have been adopted as $M = 4.0^{+1.1}_{-0.6} \times 10^6 M_\odot $ and $d=8 kpc$ \citep{EventHorizonTelescope:2022xnr,EventHorizonTelescope:2022xqj} from independent stellar dynamic observations of orbits of S0-2 star by Keck telescopes and the Very Large Telescope Interferometer (VLTI) \citep{Do:2019txf,GRAVITY2019,GRAVITY2021,GRAVITY2022,EventHorizonTelescope:2022xqj}. The observed EHT image of Sgr A* has an angular shadow diameter $d_{sh}= 48.7 \pm 7\,\mu$as and Schwarzschild shadow deviation $\delta =  -0.08^{+0.09}_{-0.09}~\text{(VLTI)},-0.04^{+0.09}_{-0.10}~\text{(Keck)}$ and is consistent with the expected appearance of a Kerr black hole \citep{EventHorizonTelescope:2022xnr,EventHorizonTelescope:2022xqj}; while no stringent comment on the inclination angle has been made, the possibility of an inclination $> 52$\textdegree \,has been disfavoured \citep{EventHorizonTelescope:2022xqj}. 

Although our goal is to place constraints on the charge of black holes, there are several caveats to this theoretical analysis, due to various underlying uncertainties, including those in the measurements of mass and distance of the central black hole. The EHT results already take into account many of these uncertainties to obtain the bounds on observables. 
For the present analysis, we consider priors on mass and distance of Sgr A*, $M \sim 4.0 \times 10^6 M_\odot $ and $d=8 kpc$ respectively, for the sake of simplification, and our method would yield constraints on the black hole parameters in exactly the same way for different mass priors. Also, while the EHT observations contain far more information related to the image of Sgr A*, to put constraints on the charges, we shall use only the EHT bounds on the two observables: angular shadow diameter $d_{sh}$ and Schwarzschild shadow deviation $\delta$ \citep{EventHorizonTelescope:2022xqj}. We fix $\theta_0=\pi/2$ considering the fact that the dependence of the shadow diameter on the inclination angle is much weaker than that on the spin in the Kerr black holes, and similarly the effect of inclination angle can be considered to be subdominant for Kerr-like black holes. Moreover, reducing the inclination angle would typically cause a reduction in the shadow size and, therefore any constraints obtained on the charges at $\theta_0=\pi/2$ can be considered as upper limits.

For estimating the angular diameter of the shadow of the Kerr-like black holes, we define the area enclosed within the shadow silhouette as \citep{Kumar:2018ple},
\begin{eqnarray}
A&=&2\int_{r_p^{-}}^{r_p^+}\left( Y(r_p) \frac{dX(r_p)}{dr_p}\right)dr_p,\label{Area}
\end{eqnarray} 
where $r_p^{\mp}$ are the prograde and retrograde spherical photon orbit (SPO) radii given respectively by the smallest and largest roots of  $\eta_{c}=0$ and $\xi_{c}(r_p^\mp)\gtrless0$ \citep{Teo:2020sey}.
Then the angular shadow diameter, for a distance $d$ from the black hole, is defined as
\citep{Bambi:2019tjh,Kumar:2020owy,Afrin:2021imp}
\begin{eqnarray}
d_{sh}=2\frac{R_a}{d}\;,\;R_a=\sqrt{A/\pi},\label{angularDiameterEq}
\end{eqnarray}  
$R_a$ being the areal shadow radius. Apart from $d$, the $d_{sh}$ also depends on the mass $M$ of the black hole.
The EHT image of Sgr A* exhibits a bright thick ring of emission with a diameter of 51.8 $\pm$ 2.3 $\mu$as -- consistent with the expectation from the black hole mass inferred from stellar dynamics \citep{EventHorizonTelescope:2022xnr} -- surrounding a brightness depression, namely the black hole shadow \citep{Afrin:2022ztr,EventHorizonTelescope:2022xnr,EventHorizonTelescope:2022xqj}. 
The diameter of the shadow  $\Tilde{d}_{metric}$\,, can measure the properties associated with the black hole metric and determine its agreement with the Kerr solution of GR for a black hole of a given angular size $\theta_g$ \citep{EventHorizonTelescope:2022xnr,EventHorizonTelescope:2022xqj}. 
\begin{figure*}[t]
\begin{center}
    \begin{tabular}{c c}
     \hspace{-0.6cm}\includegraphics[scale=0.90]{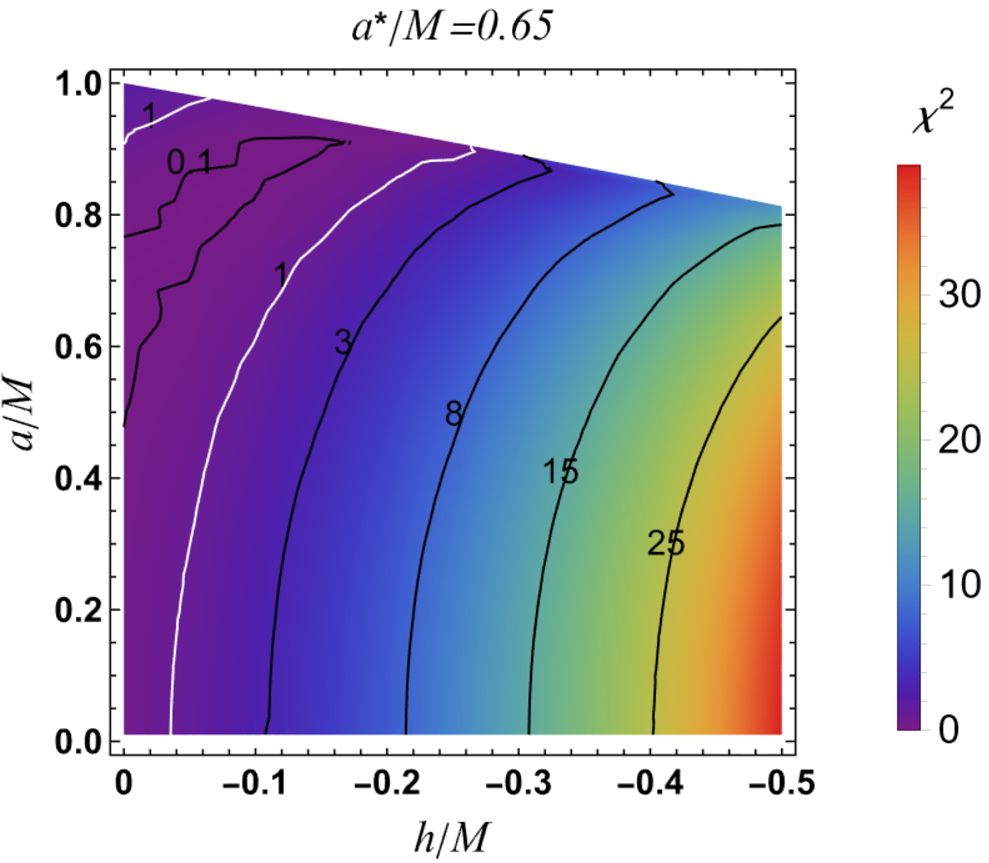}&
     \includegraphics[scale=0.90]{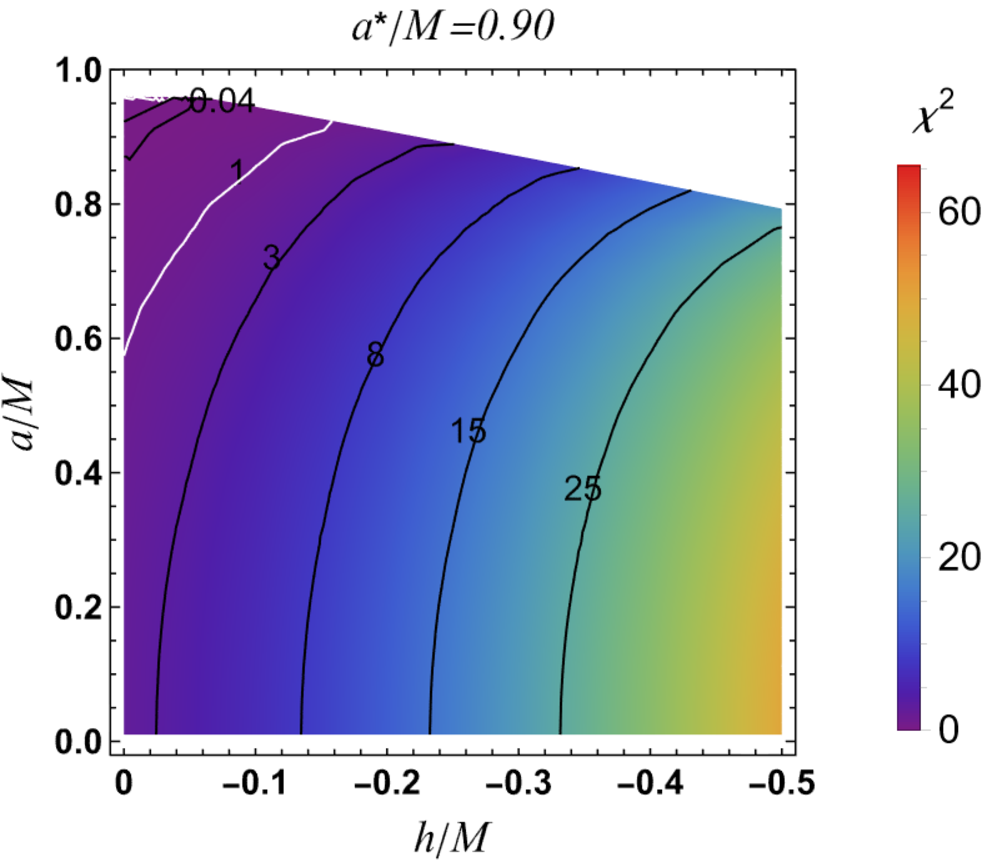}
\end{tabular}
\end{center}
	\caption{Reduced $\chi^2$ between injected Kerr shadow and modeled rotating Horndeski black hole shadow in the parameter space constrained with EHT results for Sgr A* for injected spin $a^*=0.65 M$ (left) and $0.90 M$ (right). The white region is a forbidden parameter space.}
	\label{ChiSqrFigure2}
\end{figure*}
The Schwarzschild shadow deviation ($\delta$) quantifies the deviation of the model shadow diameter ($\Tilde{d}_{metric}$) from the Schwarzschild shadow diameter $6\sqrt{3}M$ and is given by \citep{EventHorizonTelescope:2022xnr,EventHorizonTelescope:2022xqj}
\begin{equation}\label{SchwarzschildShadowDiameter}
    \delta=\frac{\Tilde{d}_{metric}}{6\sqrt{3}}-1.
\end{equation}
We take $\Tilde{d}_{metric}=2R_a$ where $R_a$ is given by Eq.~(\ref{angularDiameterEq}). The Kerr shadow diameter differs from the Schwarzschild diameter by 7.5\% as spin varies from 0 to $M$ and inclination from 0 to $\pi/2$; thus $\delta\in[-0.075,0]$  implies consistency of a model with Kerr predictions while values outside this range would clearly show discord \citep{EventHorizonTelescope:2022xqj}. Interestingly, the EHT has inferred bounds on $\delta$ \citep{EventHorizonTelescope:2022xnr,EventHorizonTelescope:2022xqj}, which comprise allowed values $\delta<-0.075 \cup \delta>0$ well outside the Kerr-accordant range $\delta\in[-0.075,0]$. This opens possibilities of testing alternative theories of gravity that predict shadows, both smaller than ($\delta<-0.075$) and larger than ($\delta>0$) Kerr. All the models that we consider -- casting shadows smaller or larger than the corresponding Kerr shadows -- are thus candidates for Sgr A*.

Using the mass and distance of Sgr A* considered by EHT, the angular diameter of the shadow is calculated for the three  Kerr-like black holes; and it is seen that for Kerr--Newman and rotating hairy black holes the $d_{sh}$ becomes smaller with increasing charges $Q$ and $l_0$ respectively, whereas for the rotating Horndeski black holes, the shadows get bigger in diameter with increasing $|h|$ (see left panels in Figure~\ref{ObservablesPlot}).
Next, we calculate the Schwarzschild shadow deviation of all the three Kerr-like black holes and find $\delta<0$ for both the Kerr--Newman and rotating hairy black holes, whereas $\delta>0$ for the rotating Horndeski black holes (see right panels in Figure~\ref{ObservablesPlot}). For  the Kerr--Newman and rotating hairy black holes, the $\delta$ decreases more than for the corresponding values of the Kerr black hole with increasing charges $Q$ and $l_0$ respectively; for the rotating Horndeski black holes, the behavior as $\delta$ increases is opposite to that for Kerr black holes, with increasing parameter $|h|$ (see right panels in Figure~\ref{ObservablesPlot}). 

It is clear from Figure~\ref{ObservablesPlot} that the EHT observations can put constraints on the charges of the Kerr-like black holes. But, since the constraints on charge parameters and spin are correlated, the placing of constraints simultaneously on both the spin and charge of the Kerr-like black holes would require further analysis in the two-parameter subspace \citep{Kumar:2020yem,Afrin:2021wlj,Afrin:2021ggx,KumarWalia:2022aop}. Modeling Sgr A* as the three Kerr-like black holes goverened by Equations~(\ref{KN_mass})-(\ref{HairyKerr_mass}), we impose the $1\sigma$ bounds $41.7\mu as\leq d_{sh} \leq 55.7\mu as$ (see Figure~\ref{AngDiamFigure}) and $-0.14\lesssim\delta\lesssim0.05$ (Keck), $-0.17\lesssim\delta\lesssim0.01$ (VLTI) (see Figure~\ref{ShadowDeviationFigure}) to find the limit on the charge.

\paragraph{Kerr--Newman black holes.} 
We note the upper limits, $Q/M\in(0, 0.90]$ \citep{EventHorizonTelescope:2021dqv} and $Q/M\in(0, 0.84]$ \citep{EventHorizonTelescope:2022xqj} respectively considering M87* and Sgr A* as Reissner--Nordstr\"{o}m black holes. From the bounds on $d_{sh}$ of Sgr A* (see Figure~\ref{AngDiamFigure}), the consistent range of charge of the Kerr--Newman black holes becomes $Q/M\in(0, 0.915]$ at $a=0$ and $Q/M\in(0, 0.8478]$ at $a/M=0.515$. Further, imposing the bounds on $\delta$ (see Figure~\ref{ShadowDeviationFigure}), the limits on charge come out to be $Q/M\in(0, 0.8915]$ (VLTI), $Q/M\in(0, 0.8328]$ (Keck) at $a=0$ and  $Q/M\in(0, 0.8255]$ (VLTI) at $a=0.5611 M$, $Q/M\in(0, 0.7174]$ (Keck) at $a=0.6990 M$. Together with the bounds on $d_{sh}$ and $\delta$ of Sgr A*, we infer $Q/M\in(0, 0.7174]$ for $a/M\in[0,1]$.
 Next, to compare with the limits imposed by M87*, we must impose the bound $\delta=-0.01\pm0.17$ \citep{EventHorizonTelescope:2019ggy,EventHorizonTelescope:2020qrl,EventHorizonTelescope:2021dqv} on the Schwarzschild deviation (see Figure~\ref{KN_M87}) with the EHT-inferred results for M87*, to find $Q/M\in(0, 0.9149]$ at $a=0$ and $Q/M\in(0, 0.8553]$ at $a=0.5176M$. Thus, $Q/M\in(0, 0.8553]$ for $a/M\in[0,1]$ is consistent with the M87* results. Hence, a more stringent upper limit can be placed on the charge of Kerr--Newman black hole with the EHT results for Sgr A* than with those for M87*.

\paragraph{Rotating Horndeski black holes.} 
From the EHT observational bounds on M87*, the inferred range of the charge parameter for the rotating Horndeski black holes has been found to be $h/M\in[-0.7564, 0)$ for $a/M\in[0.0077, 0.9353]$ \citep{Afrin:2021wlj}.
With the bounds on $d_{sh}$ of Sgr A* (see Figure~\ref{AngDiamFigure}), we obtain the limits $h/M\in[-0.3804, 0)$ at $a=0$ and $h/M\in[-0.599, 0)$ at $a/M=0.7734$. Next, putting the bounds on $\delta$ (see Figure~\ref{ShadowDeviationFigure}) we find  $h/M\in[-0.0475, 0)$ (VLTI), $h/M\in[-0.2325, 0)$ (Keck) at $a=0$ and $h/M\in[-0.2861, 0)$ (VLTI) at $a=0.8926$,  $h/M\in[-0.4582, 0)$ (Keck) at $a=0.8292$, to be observationally consistent. Together with the bounds on $d_{sh}$ and $\delta$ of Sgr A*, we infer the limit $h/M\in[-0.0475, 0)$ for $a/M\in[0,1]$.
Thus, with the EHT observational results for Sgr A*, we have placed a more stringent upper limit on the absolute charge $|h|$ of rotating Horndeski black holes than previously put with the results for M87*.

\paragraph{Rotating hairy black holes}
The EHT observational bounds of M87* set a limit $l_0/M\in[0.7122, 1)$ \citep{Afrin:2021imp}.
With the bounds on $d_{sh}$ (see Figure~\ref{AngDiamFigure}) of Sgr A*, we find that the entire parameter range $l_0/M\in(0, 1]$ and $a/M\in[0, 1]$ is consistent at current observational precision of the EHT, whereas the bounds on $\delta$ (see Figure~\ref{ShadowDeviationFigure}) place a limit $l_0/M\in[0.0696, 1)$.
Clearly, the EHT observational results for Sgr A* place more stringent lower limit on the charge of rotating hairy black holes than placed by the M87* results.
\begin{figure*}[t]
\begin{center}
    \begin{tabular}{c c}
     \hspace{-0.6cm}\includegraphics[scale=0.90]{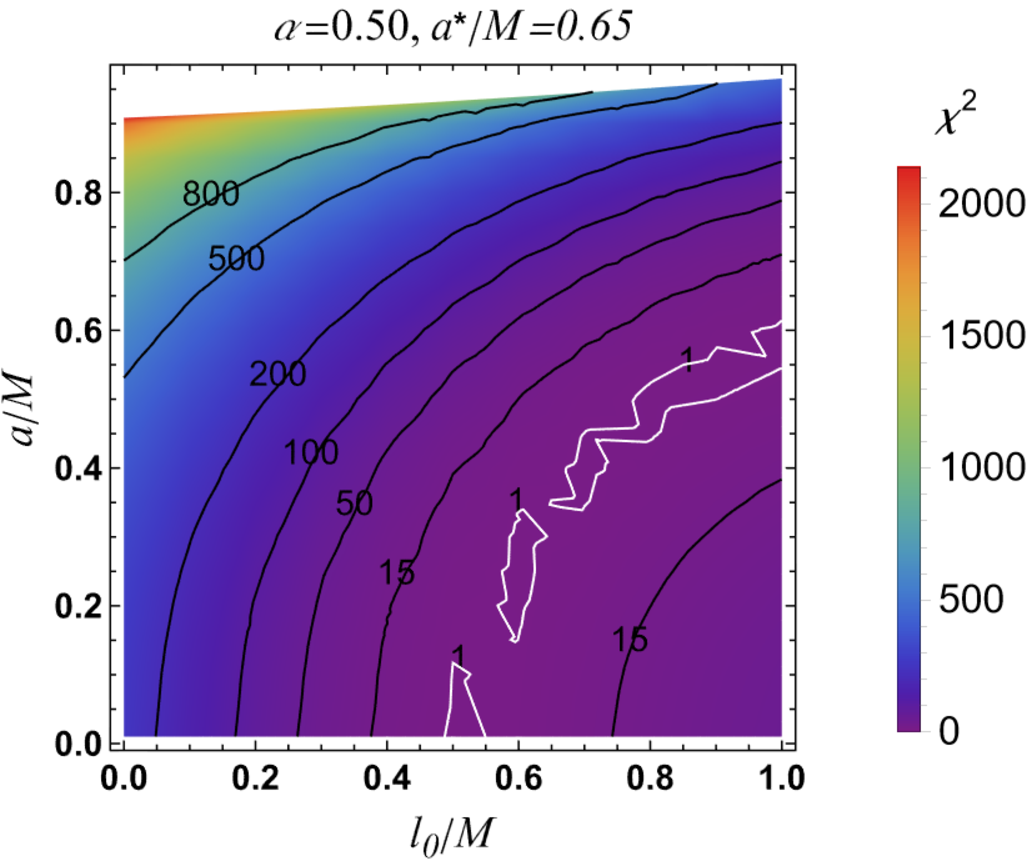}&
     \includegraphics[scale=0.90]{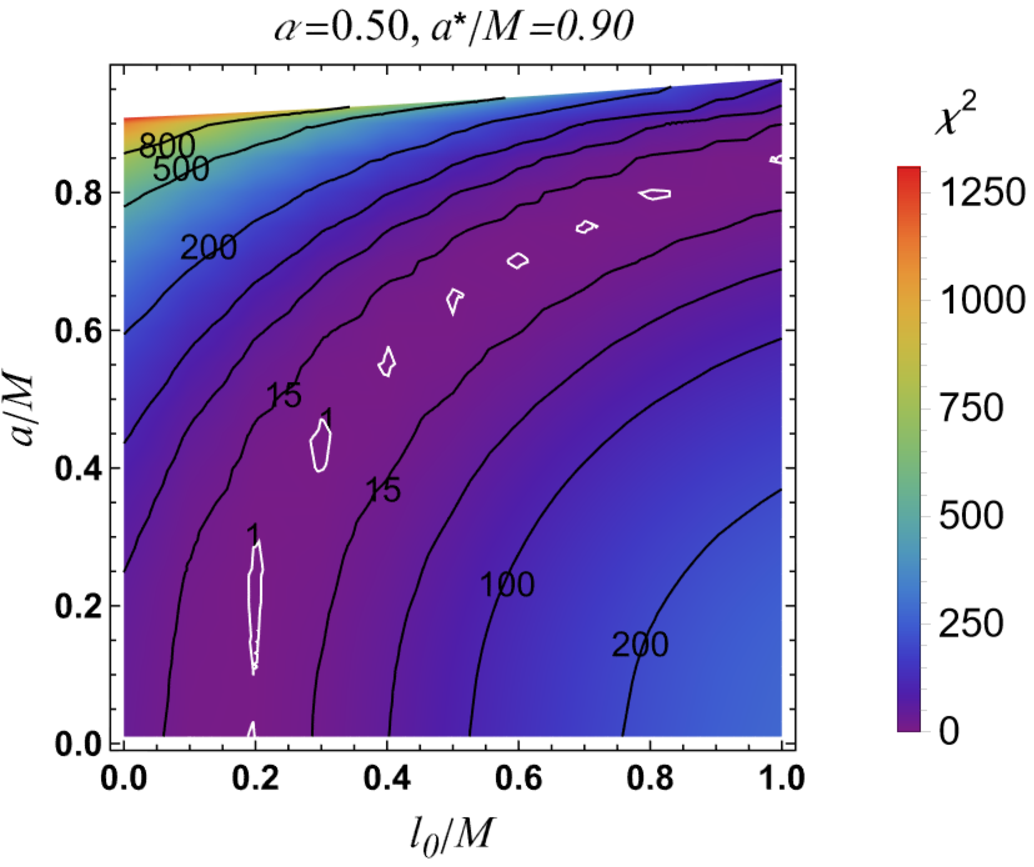}
\end{tabular}
\end{center}
	\caption{Reduced $\chi^2$ between injected Kerr shadow and modeled rotating hairy black hole shadow in the parameter space constrained with EHT results for Sgr A* for injected spin $a^*=0.65 M$ (left) and $0.90 M$ (right). The white line denotes the $\chi^2=1$ contour and  the white region is a forbidden parameter space.}
	\label{ChiSqrFigure3}
\end{figure*}
\section{Constraining with future EHT experiments}\label{sect4}
The EHT images encode signatures of various astrophysical processes besides the shadow outline, which need much better observational resolution to be disentangled from each other \citep{Lara:2021zth}. Nevertheless, this resolution is expected to be achieved with future Earth- and space-based  instruments using very long baseline interferometry, and the constraints achieved with present EHT images would most likely be superseded by new constraints, which would be partly dependent on the measurement errors as well. Therefore, we perform an analysis for future EHT experiments to demonstrate the dependence of the constraints on the error bars.

We demonstrated that the shadows cast by Kerr-like black holes show a prompt difference from those cast by Kerr black holes with varying charge (see Figure~\ref{shadow_Figure}). Besides, the various charges arising from the modifications to GR may change the shadow characteristics in ways similar to the changes caused by the spin parameter in the Kerr metric, thus causing degeneracy in the shadows. The degeneracy can be investigated to quantify the agreement between the underlying Kerr-like metric and the Kerr metric. Thus, if the future EHT observations of Sgr A* favor nonzero charges, it would suggest a scope for potential modifications of the Kerr metric within the then observational uncertainties. \citep{Ayzenberg:2018jip,Kumar:2020yem,Afrin:2021wlj}. This forms the basis of our formalism for constraining the Kerr-like metric with more precise observational capabilities.

Considering the shadows of Kerr-like black holes as models for Sgr A* and the Kerr black hole shadow as injection, we carry out a systematic bias analysis between the shadows with a reduced $\chi^2$ merit function \citep{Ayzenberg:2018jip,Kumar:2020yem,Afrin:2021wlj}, within the EHT-constrained parameter space (see Table~\ref{parameters_Table}). We utilize two shadow observables $d_{sh}$ and $\delta$, to form  the reduced $\chi^2$ function, which is minimized over the model parameter space to determine the detectability of any deviations from GR. The systematic bias analysis further allows us to investigate whether the deviation of the shadows of Kerr-like black holes from those of Kerr black holes is large enough to be observationally detectable within the observational uncertainty of future EHT observations. We compute the reduced $\chi^2$ which is defined as \citep{Ayzenberg:2018jip,Kumar:2020yem,Afrin:2021wlj}
\begin{equation}
\chi ^2(a,g,a^*)=\frac{1}{2}\sum ^2_{i=1}\bigg[\frac{\mathcal{W} ^i(a,g)- \mathcal{W} ^i_K(a^*)}{\sigma_i} \bigg]^2,\label{chi_equation}
\end{equation}
where $\mathcal{W}^i \equiv \{d_{sh}, \delta\}$ are the  black hole shadow observables, $g \equiv \{Q, h, l_0\}$ are the model charges and the measurement error is given by $$\sigma_i=\sqrt{\overline{{\mathcal{W} ^i}^2}-{\overline{\mathcal{W} ^i}}^2}.$$ Future EHT observations may have better error bars, such as $\sigma_i=2\%$ of the range of $\mathcal{W}^i$, while the present EHT measurement has an error bar of $\sim10\%$ \citep{Afrin:2021wlj,EventHorizonTelescope:2022xnr,EventHorizonTelescope:2022xqj}; the injected spin is denoted by $a^*$ and we utilize one 150 sample points ($a$, $g$) for the bias analysis. We consider the maximum possible deviation of the model from GR by fixing the inclination angle at $90$\textdegree. For fixed extrinsic parameters \{$r_0$, $\theta_0$\}, the injection depends solely on spin $a^*/M$, whereas the models depend on both spin $a/M$ and charge $g/M$. The reduced $\chi^2$ can be adopted as a measure of distinguishability between the model and injected shadows since $\chi^2\leq 1$ would signify that the model shadows are degenerate with the injected Kerr shadows and the underlying theory of gravity is indistinguishable from GR at a given observational precision, and thus constraints can be placed on the charges. On the other hand,  $\chi^2> 1$ implies the nonconformity of the model and injected shadows.

\paragraph{Kerr--Newman black holes.}
The reduced $\chi^2$ between the injected Kerr black hole shadow and model Kerr--Newman shadow, for $a^*=0.65 M \text{and} 0.90 M$, within the parameter space ($a$, $Q$) constrained with the EHT results for Sgr A* is depicted in Figure~\ref{ChiSqrFigure1}. The $\chi^2$ does not exhibit a monotonic behavior with either $Q$ or with $a$; besides, the  $\chi^2<1$ region shifts to higher values of $Q$ with an increase in injected spin $a^*$ which displays a correlation between $Q$ and injected spin $a^*$.  Further at 2\% error bar, bounds can be placed on $Q$ that would be dependent on the model $a$, namely for $a^*=0.65 M \text{and} 0.90 M$, $\chi^2<1$ in the ranges $0.1593 M\lesssim Q \lesssim 0.2587 M$ and $0.3681 M\lesssim Q \lesssim 0.3954 M$ respectively for $a=0.4M$
\paragraph{Rotating Horndeski black holes.}
To check whether the model rotating Horndeski black holes can mimic the injected Kerr black hole shadows, we have calculated the reduced $\chi^2$ between them for two fixed values of injected spins, i.e., namely $a^*=0.65 M \text{and} 0.90 M$ (see Figure~\ref{ChiSqrFigure2}). The $\chi^2$ is an increasing function of $|h|$ and is of relatively higher value for smaller values of $h$; however, $\chi^2$ decreases with $a^*$. Furthermore, the minimum $\chi^2$ contour shifts to a higher value of model spin $a$ with an increase in injected spin $a^*$, which shows a one-to-one correlation between the model and injected spins. Interestingly, $\chi^2<1$ in only a small region of the parameter space constrained earlier with the EHT results for Sgr A*, and model spin-dependent upper limits can be placed on $h$, i.e., for $a^*=0.65 M \text{and} 0.90 M$, $\chi^2<1$ in the ranges $-0.1754 M\lesssim h \lesssim 0$ and $-0.0670 M\lesssim h \lesssim 0$ respectively for $a=0.8M$. 

\paragraph{Rotating hairy black holes.}
Taking rotating hairy black holes as a model, the reduced $\chi^2$ between the model shadow and injection, within the parameter space ($a$, $l_0$) constrained with the EHT results of Sgr A*, is depicted in Figure~\ref{ChiSqrFigure3}; from this we see that  $\chi^2$ increases with increasing $a^*$. Further, $\chi^2<1$ places model spin-dependent bounds: $0.6742 M\lesssim l_0 \lesssim 0.7234 M$ and $0.2881 M\lesssim l_0 \lesssim 0.3078 M$ at $a^*=0.65 M \text{and} 0.90 M$ respectively, for $a=0.4$, such that the model shadows completely capture the injected Kerr shadows and the two are indiscernible from with EHT observations of Sgr A*. Also, the $\chi^2=1$ contour shifts to a lower $l_0$ value for higher $a^*$, hinting at a correlation between $l_0$ and $a^*$.
\section{Conclusion}\label{conclusion}
The EHT observations of the shadow of supermassive black hole Sgr A*, at the center of our galaxy the Milky Way, are an ideal and natural laboratory for testing the properties of black holes and the nature of strong-field gravity.
Together, the EHT bounds on shadow observables $d_{sh}$ and $\delta$ put a constraint on the charge of Sgr A*. For Kerr--Newman black holes the observational results for Sgr A* put upper limit on the charge, $Q/M\in(0, 0.7174]$ for $a\in[0, 1]$ whereas the bounds of M87* yield the limit $Q/M\in(0, 0.8553]$. In the case of a rotating Horndeski black hole, with Sgr A* bounds we have constrained the charge parameter to be $h/M\in[-0.0475, 0)$ for $a/M\in[0,1]$ whereas earlier studies show $h/M\in[-0.7564, 0)$ for $a/M\in[0.0077, 0.9353]$ with observational bounds inferred from the image of M87*. For the rotating hairy black holes, our analysis places bounds $l_0/M\in[0.0696, 1)$ on the charge whereas previous studies with the M87* results have reported $l_0/M\in[0.7122, 1)$.
Thus, we show, as a first, that the EHT observations of Sgr A* can place more stringent upper limits on the charges of Kerr-like black holes than those reported with EHT observation of M87*.

We conduct a chi-square analysis to assess the dependence of limits placed on charges, on the measurement error of the observational facilities. We find that as the measurement errors decrease with future Earth- as well space-based EHT imaging ventures, there is a possibility to put more stringent upper limits, as well as lower limits, on the charges of Kerr-like black holes. For example, taking two injected spins $a^*=0.65 M, 0.90 M$ the charge of Kerr--Newman black hole is constrained to be in the ranges $Q/M \in [0.1593, 0.2587]$ and $Q/M \in [0.3681, 0.3954]$  respectively for $a/M=0.4$; the charge of a rotating Horndeski black hole is constrained to be in $h/M \in [-0.1754, 0)$ and $h/M \in [0.0670, 0)$  respectively for $a/M=0.8$; whereas the charge of a rotating hairy black hole is constrained to be in $l_0/M \in [0.6742, 0.7234]$ and $l_0/M \in [0.2881, 0.3078]$  respectively for $a/M=0.4$.

Having constrained the charges, we have shown that it is difficult to rule out Kerr-like black holes from being suitable candidates for astrophysical black holes, with current and future EHT observations of Sgr A*.
\section*{Acknowledgements}
M.A. is supported by the DST-INSPIRE Fellowship, Department of Science and Technology, Govt. of India. S.G.G. thanks SERB-DST for project No. CRG/2021/005771.
\bibliography{bibfile}{}
\bibliographystyle{aasjournal}
\end{document}